\documentclass[%
two column,
reprint,%
]{revtex4-1}

\usepackage{graphicx}% Include figure files
\usepackage{dcolumn}% Align table columns on decimal point
\usepackage{bm}% bold math
\usepackage{xr}
\usepackage{amsmath}
\usepackage{color}
\usepackage[normalem]{ulem}
\usepackage{pdfpages}
\begin{document}
\markboth{\jobname}{\jobname .tex}
\preprint{}

\title{Tuning equilibration of quantum Hall edge states in graphene – role of crossed electric and magnetic fields
%Interplay of electric and magnetic field in an array of {\textit{p-n}} junction in graphene OR Quantum Hall effect in graphene 1-D tunable lateral superlattice -- incomplete mixing and LL broadening OR ....
}
\author{Sudipta Dubey}
\email[]{sudipta.tifr@gmail.com}
\affiliation{Department of Condensed
Matter Physics and Materials Science, Tata Institute of Fundamental
Research, Homi Bhabha Road, Mumbai 400005, India}
\author{Mandar M. Deshmukh}
\email[]{deshmukh@tifr.res.in}
\affiliation{Department of Condensed
Matter Physics and Materials Science, Tata Institute of Fundamental
Research, Homi Bhabha Road, Mumbai 400005, India}
\date{\today}

\begin{abstract}
We probe quantum Hall effect in a tunable 1-D lateral superlattice (SL) in graphene created using electrostatic gates. Lack of equilibration is observed along edge states formed by electrostatic gates inside the superlattice. We create strong local electric field at the interface of regions of different charge densities. Crossed electric and magnetic fields modify the wavefunction of the Landau Levels (LLs) - a phenomenon unique to graphene. In the region of copropagating electrons and holes at the interface, the electric field is high enough to modify the Landau levels resulting in increased scattering that tunes equilibration of edge states and this results in large longitudinal resistance.
\end{abstract}

%\keywords{SL, graphene, oscillations} %Use showkeys class option if keyword
\maketitle

Magnetotransport across one-dimensional superlattice (SL) had been studied in \emph{two-dimensional electron gas in semiconductor heterostructures} (2DEGS) \cite{muller_quantum_1995,tornow_even-odd_1996,ye_magnetotransport_1996,stormer_atomically_1991,endo_magnetotransport_2001,yang_zener_2002}, reporting dissipationless transport across high potential barriers \cite{muller_quantum_1995} and magnetic commensurability oscillations in longitudinal resistance \cite{ye_magnetotransport_1996}. The motivation was to study various competing length scales and energy scales between tunable SL potential and quantum Hall system. Graphene offers the advantage of large cyclotron gap allowing quantum Hall effect to be observed at room temperature \cite{castro_neto_electronic_2009,das_sarma_electronic_2011,goerbig_electronic_2011}. Substrate induced SL in graphene in the presence of magnetic field led to the experimental observation of Hofstadter butterfly physics \cite{dean_hofstadters_2013,ponomarenko_cloning_2013,hunt_massive_2013}. The ability to create abrupt ($\sim$ 10~nm) tunable barriers in graphene allows new aspects to be explored. In addition, new physics, due to the role of crossed electric and magnetic field, that cannot be seen in conventional 2DEGS can be studied in SL structures based on graphene.

In this letter, we study magneto transport in an electrostatically defined 1D lateral SL in graphene \cite{dubey_tunable_2013}. In our device we apply a perpendicular magnetic field and periodically modulate the charge carrier density in adjacent ``ribbons" of graphene, tuning from an array of {\textit{p-p'}} (or {\textit{n-n'}}) to an array of {\textit{p-n'}} junctions. Changing the magnetic field allows us to vary $l_B$ relative to $\lambda$; and changing the gate voltage allows us to tune the SL potential strength relative to LL spacing. The relative abruptness, bipolarity of charge carriers, large modulation and unequally spaced LLs distinguishes the present work from the previous work on 1D SL using 2DEGS systems \cite{muller_quantum_1995,tornow_even-odd_1996,ye_magnetotransport_1996,stormer_atomically_1991,kawamura_quantum_2001}.

Apart from the length scales, we also study the energy scales involved. The competition between SL amplitude ($V_0$) and LL spacing ($\hbar\omega_c$, where $\hbar=h/2\pi$, h being the Planck's constant, and $\omega_c$ is the cyclotron frequency) gives rise to three regimes. When $V_0 >> \hbar\omega_c$, SL effect dominates giving rise to extra Dirac points \cite{killi_graphene:_2012}. In the other extreme when $V_0 << \hbar\omega_c$, quantum Hall effect in graphene is restored \cite{killi_graphene:_2012}. However, the situation is more complex and little explored when $V_0$ and $\hbar\omega_c$ have comparable contribution, and we have experimentally probed this regime in graphene.

The goal of our work is to extend quantum Hall studies beyond single top-gate in graphene \cite{ozyilmaz_electronic_2007,ki_quantum_2009,gu_collapse_2011}. Our work is the first experimental report on magnetotransport in multiple top-gates on graphene and we probe the physics of equilibration along the narrow region in graphene defined electrostatically. Our main observation is that when $V_0$ is comparatively small in the unipolar region, the edge states do not equilibrate along this narrow region defined electrostatically. The extent of equilibration can be tuned in the bipolar region where the electric field is relatively large. In this regime, electric field significantly modifies the Landau level wavefunctions, increasing scattering, which is reflected in increased equilibration and large longitudinal resistance.

\begin{figure}
\includegraphics[width=85mm]{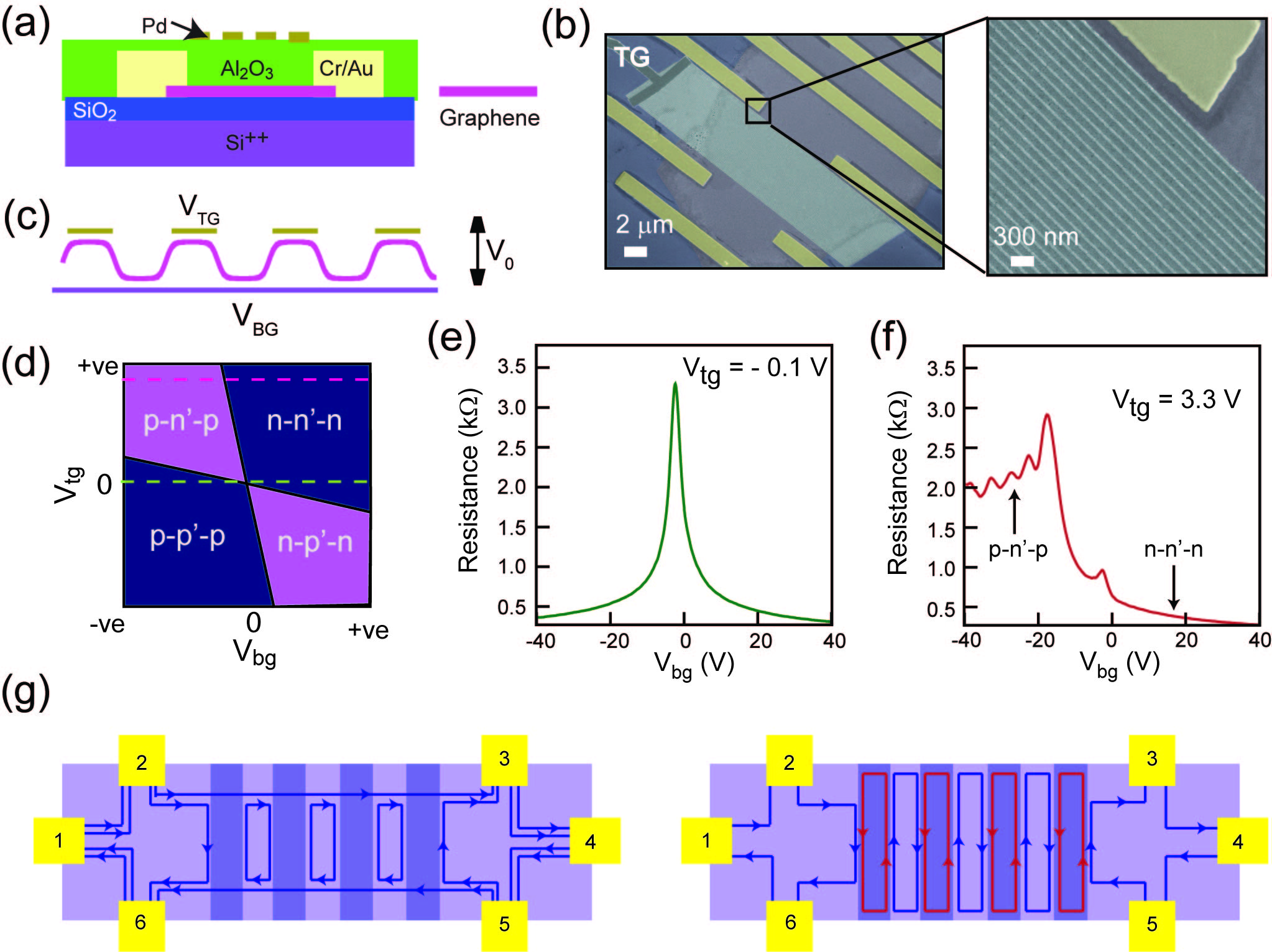}
\caption{\label{fig:figure1} Device geometry and different regimes of edge state transport. (a) Schematic of a device.
(b) False colored scanning electron microscope image of a device with a zoomed-in image of the finger like top gates.
(c) Depiction of the periodic 1D potential and the influence of the two gates in BG and TG regions. (d) Parameter space of $V_{tg}$ and $V_{bg}$  showing the different type of charge carriers in adjacent regions.
(e) Line plot of resistance as function of $V_{bg}$ when $V_{tg}$ is biased in charge neutral region at 0~T and 2~K.
(f) Line plot of resistance at 0~T and 2~K when TG regions are n-doped.
(g) Schematic depicting edge state transport in the unipolar region with $\nu_{bg}$ greater than $\nu_{tg}$, and the bipolar region. The number of top-gates is four in this schematic. The BG region is denoted in purple and the TG region is shaded in blue.}
\end{figure}

We create a 1D tunable SL, of period $\lambda$, by fabricating an array of thin finger gates on graphene. The schematic of a device is shown in Figure~\ref{fig:figure1}(a), and Figure~\ref{fig:figure1}(b) shows false colored scanning electron microscope image (details of fabrication in Section~I of Supplemental Material). The geometric width of each top-gates is $\sim$ 30~nm and they have a period of $\lambda$ $=$ 150~nm. The effective electrostatic width of the top-gates felt by the charge carriers in graphene is larger due to the finite thickness of the top-gate dielectric \cite{dubey_tunable_2013} (details in Section~VII of Supplemental Material).

In our device, graphene consists of two alternating regions - one where the charge carrier density is controlled only by the back-gate (BG region); and the other where the charge carrier density is controlled by both the top-gate and the back-gate (TG region). The difference in charge carrier density between BG and TG regions gives rise to a SL whose amplitude ($V_0$) is controlled by $V_{bg}$ and $V_{tg}$. ($V_0 = \sqrt{\pi}\hbar v_F (sgn(C_{bg}V_{bg})\sqrt{\frac{|C_{bg}V_{bg}|}{e}} -sgn(C_{tg}V_{tg}+C_{bg}V_{bg})\sqrt{\frac{|C_{tg}V_{tg}+C_{bg}V_{bg}|}{e}})$\cite{dubey_tunable_2013}, where $V_{tg}$ ($V_{bg}$) is the top-gate (back-gate) voltage, $C_{tg}$ ($C_{bg}$) is the capacitance per unit area of top-gate (back-gate), $e$ is the electronic charge and $v_F$ is the Fermi velocity.) (Details of calculation and plot of $V_0$ as a function of $V_{bg}$ and $V_{tg}$ is in Section~II of Supplemental Material.)

We measure zero-bias \emph{four-probe} longitudinal resistance ($R_{xx}$) while varying gate voltages at different magnetic fields ($B$) at a temperature of 2~K. The charge neutral point is at $V_{tg}$ $=$ -0.1~V and $V_{bg}$ $=$ -2~V (Figure~\ref{fig:figure1}(e)) suggesting low unintentional doping. The mean free path in our device is $\sim$ 70~nm and phase coherence length is $\sim$ 600~nm at 2~K \cite{tikhonenko_weak_2008}. As $B$ increases, the magnetic length ($l_B=\sqrt{\frac{\hbar}{eB}} $) decreases and the charge carriers encounter smaller periods of SL until they are confined within a single BG or TG region. Well resolved LLs start to appear only beyond 2~T (see Supplemental Material Section~III for complete LL fan diagram). In this work we look at quantum Hall effect in 1D SL at $\frac{\lambda}{l_B}=$ 22 and vary $V_0$ upto $\sim$ 375~meV.

The resistance depends on the filling factor $\nu=nh/Be$ in the adjacent regions \cite{abanin_quantized_2007,williams_quantum_2007,ozyilmaz_electronic_2007,ki_quantum_2009}. At a constant magnetic field B, the filling factor in the adjacent regions is determined by $V_{bg}$ and $V_{tg}$. Depending on $V_{bg}$ and $V_{tg}$ at a given $B$, we have same type of edge states in the unipolar region, and in the bipolar region, electron and hole edge states co-propagate along the junction (Figure~\ref{fig:figure1}(g)). In this device, we have 37 top-gates. So, if the edge states are formed under all the top-gates and they equilibrate along all the edges of the top-gates, then the \emph{four-probe} resistance ($R_{xx}$) plateau is given by
\[
 R_{xx} = \left\{
  \begin{array}{l l l}
    \frac{h}{e^2}\frac{N||\nu_{tg}|-|\nu_{bg}||}{|\nu_{tg}||\nu{_{bg}}|} & \quad \nu_{tg}\nu_{bg}>0 & \quad \text{(I)}\\
    \frac{h}{e^2}\frac{N(|\nu_{tg}|+|\nu_{bg}|)}{|\nu_{tg}||\nu_{bg}|} & \quad \nu_{tg}\nu_{bg}<0 & \quad \text{(II)}
  \end{array} \right.\]
where $N$ is the number of top-gates which is 37 in our device.	(Landau-B\"{u}ttiker formalism to obtain $R_{xx}$ for multiple top-gates in Section~V of Supplemental Material.)

Figure~\ref{fig:figure2}(a) shows the colorscale plot of zero bias \emph{four-probe} resistance as a function of $V_{tg}$ and $V_{bg}$ at 14~T. We observe diamond shaped regions in the parameter space that represent integer filling factors in adjacent regions set by $V_{tg}$ and $V_{bg}$. The filling factors in the two alternating regions are indicated as ($\nu_{tg}$,$\nu_{bg}$). Figure~\ref{fig:figure2}(b, c) show line slices of $R_{xx}$ as function of $\nu_{bg}$ at $\nu_{tg} =$ 2, 6. (Line slices at $\nu_{tg} =$ 10 and 14 are presented in Section~IV of Supplemental Material.) We do not observe large resistance as predicted by Equation I and II. The $R_{xx}$ is relatively high, and does not show a plateau, in the bipolar regime, but not as high as predicted by Equation~II.

In the line slices (Figure~\ref{fig:figure2}(b, c)), the green curve is the experimental data. The black dashed lines correspond to the calculated plateau for $N$ $=$ 1 in Equation I and II. We find that in the unipolar regime, denoted by the blue region in Figure~\ref{fig:figure1}(d), the potential $V_0$ is small and the plateaus are well described by $R_{xx}=\frac{h}{e^2}\frac{||\nu_{tg}|-|\nu_{bg}||}{|\nu_{tg}||\nu_{bg}|}$; there is good agreement between measured experimental data and expected plateau values for a single top-gate in Figure~\ref{fig:figure2}(b,c).

Our experimental observation show plateaus corresponding to a single top-gate and not 37 top-gate, and so we do not have equilibration in our device inside the superlattice. The equilibration occurs only at the extreme edge of the top-gates near the voltage probes as illustrated in the schematic in Figure~\ref{fig:figure2}(d). In the schematic, the yellow probes are the real probes and the gray probes are the virtual probes denoting equilibration at that edge. The virtual voltage probes are used to calculate resistance using Landauer-B\"{u}ttiker formalism \cite{buttiker_absence_1988}. One possible reason for the lack of equilibration is that the edge states under the top-gate are defined electrostatically, where the potential varies smoothly due to the finite thickness of the top-gate dielectric \cite{amet_selective_2014}. Equilibration requires inter edge state scattering or ohmic contacts so that all the edge states are at the same chemical potential, and this does not happen due to the short length along the physical edge of graphene \cite{kane_contacts_1995,alphenaar_influence_1991,ensslin_edge_2003,haug_edge-state_1993}.

\begin{figure}
\includegraphics[width=85mm]{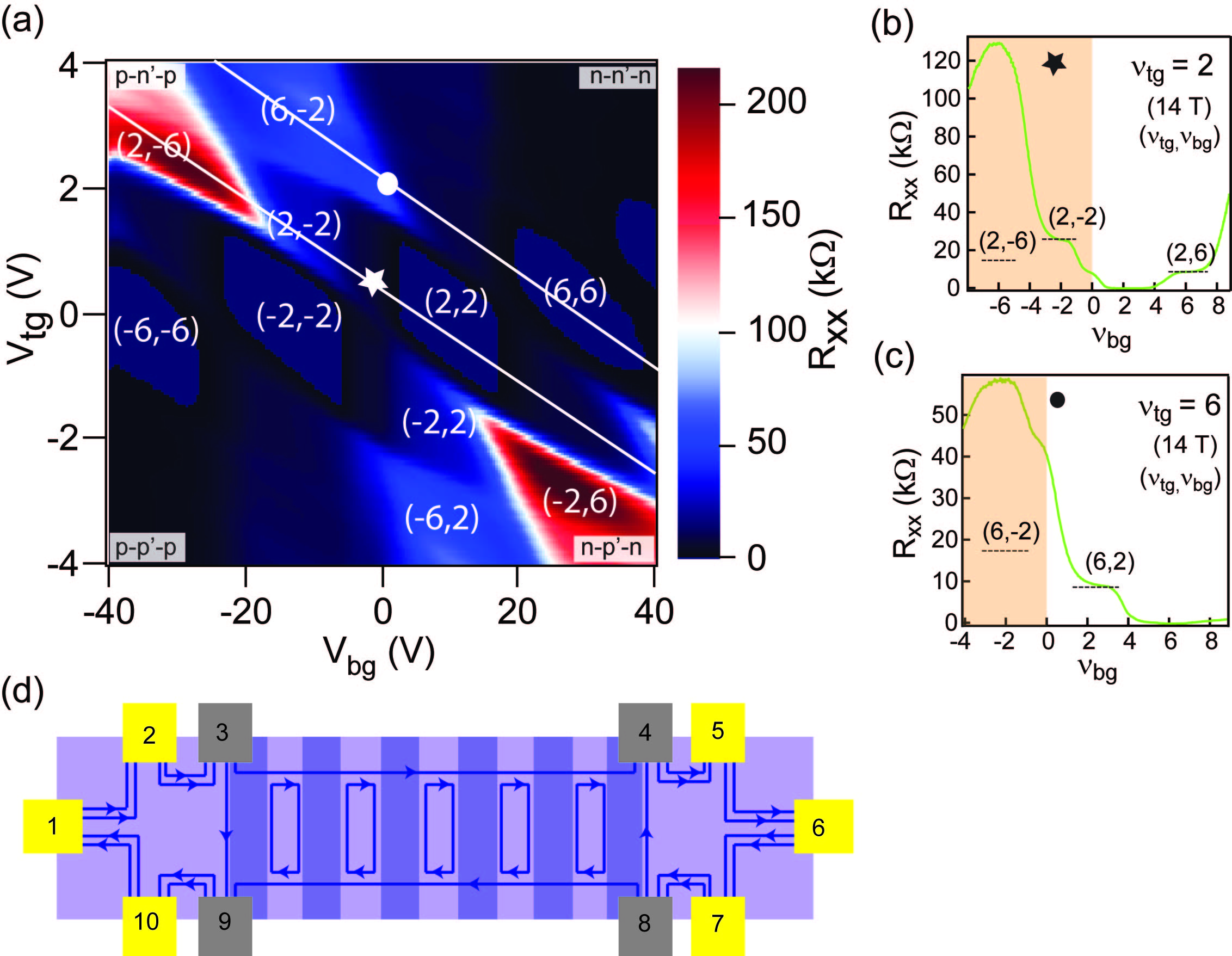}
\caption{\label{fig:figure2} Longitudinal resistance ($R_{xx}$) at $B$ $=$ 14~T when LLs are well resolved in three different case of edge state transport. (a) $R_{xx}$ as function of $V_{tg}$ and $V_{bg}$ at 14~T at a temperature of 2~K. The white lines denote line of constant $\nu_{tg}$. $R_{xx}$ as function of $\nu_{bg}$ for (b) $\nu_{tg} =$ 2 and (c) $\nu_{tg} =$ 6. The green curves in (b), (c) denote experimental value with the black dashed lines representing the resistance plateau in case of equilibration of edge states for a single top-gate. (($\nu_{tg}$,$\nu_{bg}$) values are denoted within the parenthesis.) (d) Schematic depicting equilibration along the extreme edges in unipolar region. The yellow probes are the real contacts and the gray probes are the virtual voltage probes.}
\end{figure}

In the bipolar region, plateau coincides with that expected for a single top-gate when $\nu_{tg} =$ 2(-2) and $\nu_{bg} =$ -2(2). When we have one edge state for both electrons and holes circulating in adjacent regions, the resistance plateau is seen at $h/e^2$ (25.8~k$\Omega$). The plateau at ($\nu_{tg}$,$\nu_{bg}$) $=$ (2,-2) results not only because the $E$ is low in this state compared to any other state in the bipolar region but also because this is a special state where the charge carriers belong to the same LL ($n$ $=$ 0) and that the LL gap is maximum between $n$ $=$ 0 and $n$ $=$ 1 LL. Narrow TG and BG regions inside the superlattice leads to leaky barrier and transmission of edge states resulting in a plateau corresponding to a single top-gate \cite{muller_quantum_1995}.

However, when neighboring regions have other edge states, that is, $(\nu_{tg}\times \nu_{bg}) < 0$ and $|\nu_{tg}|$ or $|\nu_{bg}|$ is greater than 2, we find resistance significantly larger than $h/e^2$ (for example, (2,-6) in Figure~\ref{fig:figure2}(b), (6,-2) in Figure~\ref{fig:figure2}(c)); a feature not generally seen in single {\textit{p-n'-p}} junction in the quantum Hall state. The feature of plateaus in $R_{xx}$ corresponding to equilibration for a single top-gate in the unipolar region and deviation from this picture in the bipolar region, is quite robust and is also observed in another device with 35 top-gates (details in Section~X of Supplemental Material).

\begin{figure}
\includegraphics[width=90mm]{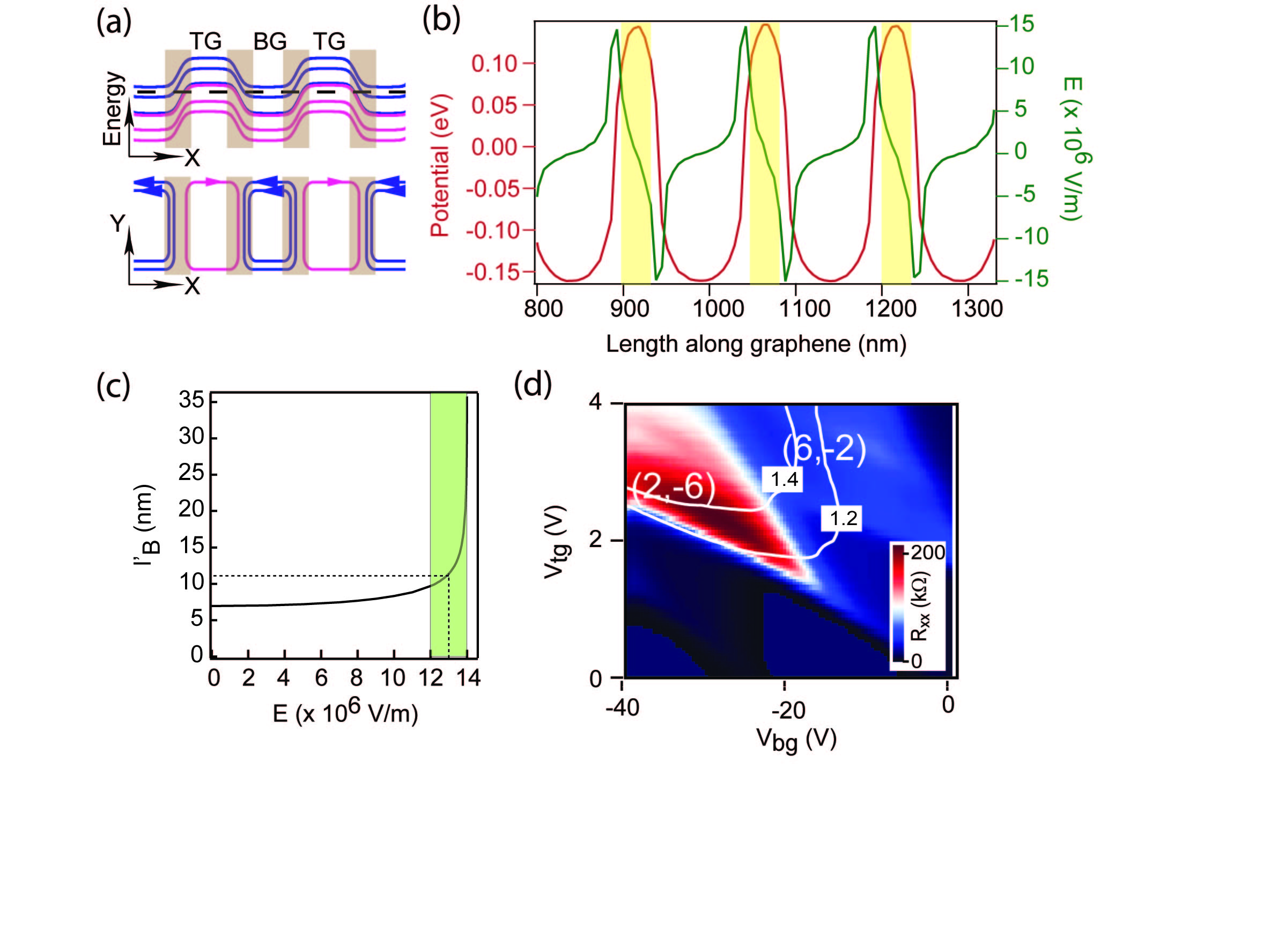}
\caption{\label{fig:figure4} Effect of $E$ on equilibration (a) Variation of LLs along the length of device when ($\nu_{tg}$,$\nu_{bg}$) $=$ (-2,6). The brown region in the schematic denotes the junction between TG and BG region where $E$ exists. (b) Spatial variation of potential and $E$ along the length of graphene in bipolar region ($V_{bg}$ $=$ -30~V and $V_{tg}$ $=$ 3~V). Yellow regions denote geometric position of top-gates. (c) Effective magnetic length as function of $E$ at $B$ $=$ 14~T. (d) Contour of maximum $E$ at 1.2$\times$10$^7$~V$/$m and 1.4$\times$10$^7$~V$/$m as function of V$_{bg}$ and V$_{tg}$ overlaid on measured R$_{xx}$ in bipolar regime at 14~T.}
\end{figure}

The large resistance with maximum $R_{xx} \sim$ 200~k$\Omega$ at 14~T, is seen along the diagonal direction in Figure~\ref{fig:figure2}(a); it is precisely in this diagonal direction of the parameter space that $V_0$ increases. In the bipolar region, though the value of $R_{xx}$ is higher than that for a single top-gate, it is not 37 times the value for a single top-gate, and thus there is no full equilibration. In this region the array of top-gates does not behave as a single top-gate. The resistance value depends on the gate voltages and the magnetic field, implying that the equilibration can be modified by electric field.

We now try to understand the reason for this observation and note that $V_0$ is relatively large in this region. Figure~\ref{fig:figure4}(a) shows a LL diagram along the length of the device when $(\nu_{tg},\nu_{bg}) = (-2,6)$; where we have co-propagating electron and hole edge states at the junction. At the interface of these regions (shaded brown in Figure~\ref{fig:figure4}(a)) there is an electric field ($E$), due to the SL, and it has a significant effect on the LL wavefunctions.

To get an accurate idea of the magnitude of $E$ in our devices we performed numerical simulation of the electrostatics using finite element method. At a given $V_{tg}$ and $V_{bg}$, the charge carrier density induced along the length of graphene is calculated, from which potential and $E$ is obtained. (Details of calculation in Section~VII of Supplemental Material.) Spatial variation of potential and $E$ along length of graphene in bipolar region is shown in Figure~\ref{fig:figure4}(b). At a given $V_{tg}$ and $V_{bg}$, the maximum $E$ is obtained which is higher in the bipolar region compared to the unipolar region.

We note that in the bipolar regime, $V_0$ created is much larger than $\hbar \omega_C$. Large $V_0$ leads to large $E$ in the region between BG and TG region which modifies the LLs locally. It has been shown by Lukose \emph{et al.} \cite{lukose_novel_2007} and later extended by Gu \emph{et al.} \cite{gu_collapse_2011} for the case of a top-gate geometry that the LL spectrum and the wavefunctions in crossed $E$ and $B$ are fundamentally modified in graphene, an aspect that is not observable in conventional 2DEGS semiconductors. LL wavefunction is modified in two ways. Effective magnetic length in the presence of $E$ can be written as $l'_B = l_B/(1-(E/v_FB)^2)^{1/4}$ \cite{lukose_novel_2007}. $l'_B$ increases with increasing $E$ and rises rapidly when $E$ approaches $v_F B$ as seen in Figure~\ref{fig:figure4}(c). Secondly, the $E$ mixes the Landau levels.

Contours of maximum $E$ at 1.2$\times$10$^7$~V$/$m and 1.4$\times$10$^7$~V$/$m are overlaid on the measured data at 14~T as shown in Figure~\ref{fig:figure4}(d). Figure~\ref{fig:figure4}(d) shows that the contours lie along the region where we have high resistance state and departure from the value for a single top-gate.  We argue that in our device geometry, $E$ created is high enough to modify LLs which is reflected in charge transport measurements. The spatial extent of the wavefunction ($l'_B$) increases with increasing $E$ and approaches the width of TG (or BG) region. This leads to increased spatial overlap of the wavefunction within TG (or BG) region of the superlattice resulting in enhanced scattering. Secondly, as function of $E$, there is LL mixing \cite{lukose_novel_2007} which results in a non zero matrix element essential to cause scattering and equilibration. Let us now examine the various lengthscales of our system that support this scenario. From Figure~\ref{fig:figure4}(b) we find that the effective electrostatic width of the top-gate for this configuration is $\sim$50~nm and the extent of the region with high electric field (between 1.2 - 1.4$\times$10$^7$~V$/$m) is $\sim$15~nm. In addition,  the magnetic length in the presence of transverse electric in this region is $\sim$11~nm (see Figure~\ref{fig:figure4}(c)). If one considers the scenario of (6,-2) state, we find that with three edge states with width ~11~nm will have significant overlap with two adjacent regions and equilibration will be enhanced due to scattering.

In the presence of high electric field, the resistance depends on $E$ and $B$ applied. For example, resistance values at ($\nu_{tg}$,$\nu_{bg}$) $=$ (6,-2) and (2,-6) at 14~T are different as the maximum $E$ is different in the two regions (Plot of maximum $E$ as a function of gate voltages in Section~VII of Supplemental Material) --- this strongly suggests electric field tuning of equilibration.

We think disorder does not play an important role. Because, in our sample, the disorder potential, estimated from the FWHM of the Dirac peak, is 71~meV \cite{dubey_tunable_2013}, and thus is smaller than $V_0$ in the bipolar region. So, the tuning of equilibration of edge states in our sample, which is seen at higher $V_0$, is due to the $E$ at the interface of TG and BG regions. However, recent work of Kumada et al suggests that the disorder along the length of p-n junction could play an important role in equilibration \cite{kumada_shot_2015,long_disorder-induced_2008}. Further experimental and theoretical studies need to be carried out to probe the role of disorder in periodically modulated structures.

Details of LL modification in unipolar region at lower $B$ of 3.5~T in Section~IX of Supplemental Material.

Our experiments with tunable superlattices suggest that tuning of the equilibration of edge states  in graphene can be done using $E$ at interfaces, which cannot be realized in conventional 2DEGS. In addition, the nature of the state that emerges after the collapse of the LLs is little understood and possibility of existence of correlations has been speculated \cite{lukose_novel_2007,carmier_semiclassical_2011}. The close proximity of co-propagating electron and hole edge states can be used to construct large class of topological states \cite{sagi_non-abelian_2014} and also offers an opportunity to study excitonic effects, this has been recently explored in bilayer quantum Hall systems \cite{nandi_exciton_2012}. There have been predictions of correlated states in $\nu$ $=$ 0 LL of graphene and LL mixing can enable exploration of such phases \cite{kharitonov_phase_2012}.

We thank Marcin Mucha-Kruczy\'nski, G.Baskaran, R.Shankar, Jainendra Jain, Vibhor Singh, Shamashis Sengupta and K.Sengupta for discussions and comments on the manuscript. We acknowledge Swarnajayanthi Fellowship of Department of Science and Technology and Department of Atomic Energy of Government of India for support.

%\bibliography{Quantum_Hall}




\section*{Supplementary material}

**Supplemental Material: Quantum Hall effect in tunable 1-D lateral superlattice in graphene – role of crossed electric and magnetic fields**

Sudipta Dubey[1, a)] and Mandar M Deshmukh[1, b)]

*Department of Condensed Matter Physics and Materials Science,*

*Tata Institute of Fundamental Research, Homi Bhabha Road, Mumbai 400005,*

*India*

---

a)sudipta.tifr@gmail.com

b)deshmukh@tifr.res.in

## I. DEVICE FABRICATION

Graphene is mechanically exfoliated on degenerately doped silicon (Si$^{++}$) substrate having 300 nm of silicon dioxide. Si$^{++}$ acts as global back-gate, that is, it tunes the charge carrier density of the entire graphene flake. Standard e-beam lithography is used to pattern Cr/Au contacts in Hall bar geometry. The device is then thermally annealed to remove residual PMMA from graphene. 3 nm of Al$_2$O$_3$ is then deposited using e-beam lithography which acts as seed layer for subsequent deposition of 20 nm Al$_2$O$_3$ using atomic layer deposition. Al$_2$O$_3$ acts as top-gate dielectric above which narrow ($\sim$ 27 nm wide), periodic fingers of Pd are patterned using e-beam lithography. All the fingers of top-gate are connected to one electrode outside the graphene flake so that all the fingers are at the same potential. The period of these fingers is 150 nm.

## II. VARIATION OF AMPLITUDE OF SUPERLATTICE POTENTIAL WITH BACK-GATE AND TOP-GATE VOLTAGE

In our device there are two alternating regions in graphene - one without a top-gate (BG region), where the charge carrier density is controlled only by the back-gate; and the other under a top-gate (TG region), where the charge carrier density is controlled by the back-gate and the top-gate. The amplitude of superlattice potential ($V_0$) created is given by the difference of doping in the BG and TG region. In case of no unintentional doping, the number density of charge carriers in the BG region is given as $n_{bg}=\frac{C_{bg}V_{bg}}{e}$; and in the TG region, the net number density of charge carriers is the algebraic sum of the density of carriers induced by the two gates $n_{tg}=\frac{C_{tg}V_{tg}+C_{bg}V_{bg}}{e}$; where $C_{bg}$ ($C_{tg}$) is the capacitance per unit area between graphene and back-gate (top-gate) and $V_{bg}$ ($V_{tg}$) is the back-gate (top-gate) voltage. Using parallel plate capacitor geometry $C_{bg}$ is given by $\epsilon_0\epsilon_r/d$, where $\epsilon_0 = 8.85\times10^{12}$ F/m, $\epsilon_r = 3.9$ and $d = 300$ nm. $C_{tg}$ is obtained from the slope of charge neutral line in the plot of resistance as function of $V_{bg}$ and $V_{tg}$ at zero magnetic field (Figure S1(a)). From the slope, we get $C_{tg} = 18\times C_{bg}$. In case of graphene, due to its linear dispersion, $E_F = \sqrt{\pi}\hbar v_F\sqrt{n}$, where $v_F$ is the Fermi velocity and $E_F$ is the Fermi energy, with all

energies measured relative to the charge neutrality point[1]. Thus $V_0$ is given by

$$V_0 = \sqrt{\pi}\hbar v_F \left( sgn(C_{bg}V_{bg})\sqrt{\frac{|C_{bg}V_{bg}|}{e}} - sgn(C_{tg}V_{tg} + C_{bg}V_{bg})\sqrt{\frac{|C_{tg}V_{tg} + C_{bg}V_{bg}|}{e}} \right) \quad (1)$$

Magnitude of $V_0$ is referred as $V_0$ since it represents the amplitude of superlattice potential. Figure S1(b) shows the variation of $V_0$ with $V_{bg}$ and $V_{tg}$.

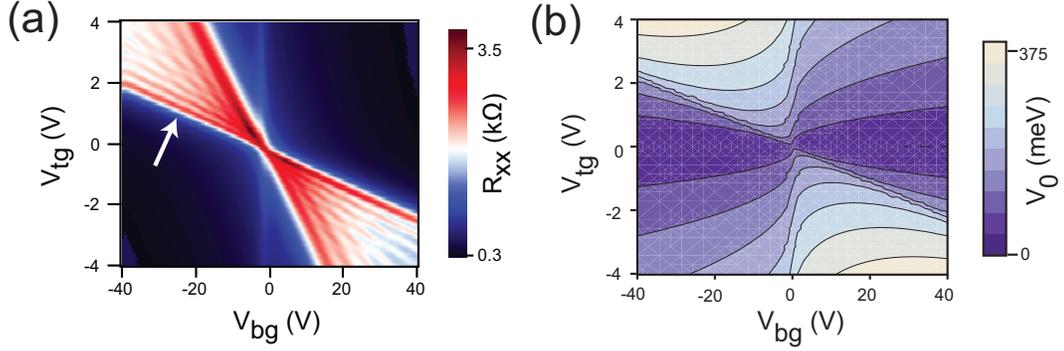

FIG. S1. Amplitude of superlattice potential (a) Resistance as function of $V_{tg}$ and $V_{bg}$ at zero magnetic field at temperature of 2 K. White arrow points to the charge neutral line. (b) Contourplot of $V_0$ as a function of $V_{bg}$ and $V_{tg}$. $V_0$ is higher in the bipolar region where there is a series of p-n' junctions compare to the unipolar region.

## III.  QUANTUM HALL IN THE ABSENCE OF SUPERLATTICE POTENTIAL

Figure S2(a) shows the magnitude of Hall resistance as function of $V_{bg}$ and magnetic field (B) when $V_{tg}$ is biased at charge neutral point ($V_{tg} = 0.1$). The fan diagram corresponds to monolayer graphene. Figure S2(b) is a line plot at 10 T showing plateau at $\nu = 2$ ($\equiv$ 12.9 k$\Omega$), $\nu = 6$ ($\equiv$ 4.3 k$\Omega$) and $\nu = 10$ ($\equiv$ 2.5 k$\Omega$), where $\nu$ is filling factor. We can observe the Landau levels from 2 T.

## IV.  OBSERVED RESISTANCE FOR DIFFERENT FILLING FACTORS IN TG AND BG REGION

The filling factor is given by $\nu = nh/Be$, where $n$ is the charge carrier density, $h$ is the Planck's constant, $B$ is the magnetic field and $e$ is the electronic charge. For constant $B$,

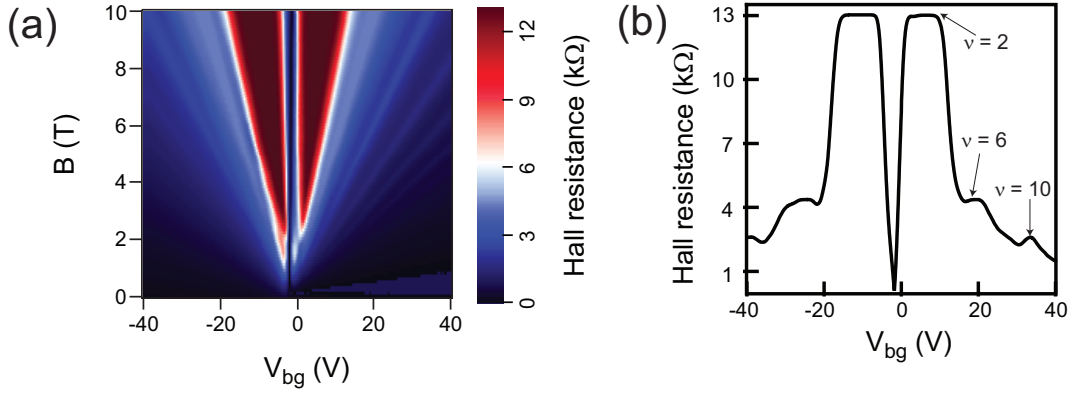

FIG. S2. (a) Magnitude of Hall resistance as a function of $V_{bg}$ and $B$ (b) Line plot of magnitude of Hall resistance as function of $V_{bg}$ at 10 T showing plateaus at $\nu = \pm\, 2, \pm\, 6, +\, 10$.

filling factor in BG region ($\nu_{bg}$) is controlled by $V_{bg}$, and filling factor in TG region ($\nu_{tg}$) is controlled by both $V_{tg}$ and $V_{bg}$.

Line slices showing longitudinal resistance ($R_{xx}$) as function of $\nu_{bg}$ for $\nu_{tg} = 2$ and 6 are shown in Figure 2 in main text. For $\nu_{tg} = 10$ and 14, we measured $R_{xx}$ as function of $V_{tg}$ and $V_{bg}$ (Figure S3(a)) at a lower field of 8 T. Figure S3(b,c) shows $R_{xx}$ as function of $\nu_{bg}$ at $\nu_{tg} = 10$ and 14. In the line slices (Figure S3(b, c)), the green curve is the experimental data with the black dashed line marking the calculated plateau in case of edge state equilibration for a single top-gate. Plateaus coincide with the black dashed lines in the unipolar region but deviates from them in the bipolar region.

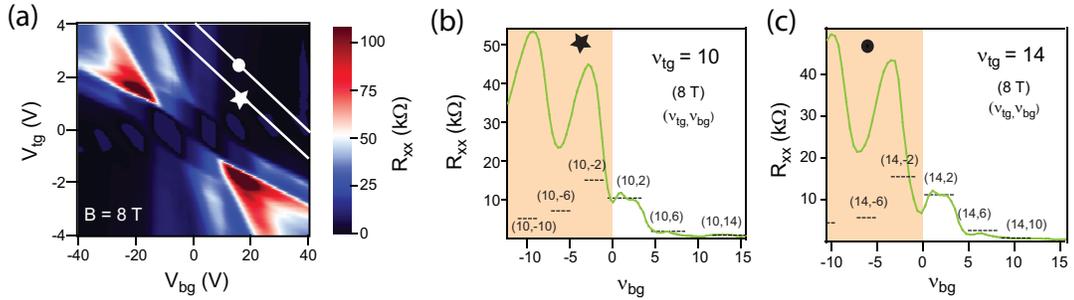

FIG. S3. (a) Longitudinal resistance ($R_{xx}$) as function of $V_{tg}$ and $V_{bg}$ at 8 T. $R_{xx}$ as function of $\nu_{bg}$ for (b) $\nu_{tg} = 10$ and (c) $\nu_{tg} = 14$. The green curves in (b), (c) denote experimental value with the black dashed lines representing the resistance plateau in case of equilibration of edge states for a single top-gate. (($\nu_{tg},\nu_{bg}$) values are denoted within the parenthesis.)

## V. RESISTANCE IN CASE OF EQUILIBRATION OF EDGE STATES

The schematic of our device geometry is shown in Figure S4. The yellow probes denote real contacts used to measure the resistances. The TG regions are denoted in blue and BG regions in purple.

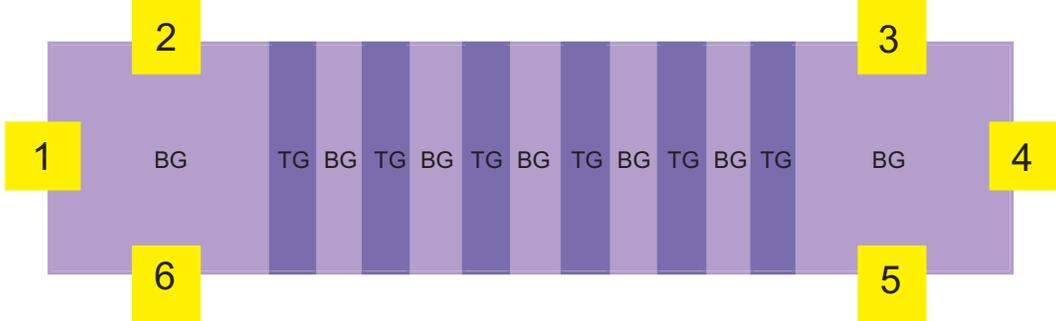

FIG. S4. Schematic showing our device geometry with multiple top-gates. In this schematic, six top-gates are shown. Yellow probes denote real contacts. Longitudinal resistance is measured between voltage terminal 2 and 3, and Hall resistance between voltage terminal 2 and 6.

We now describe the resistance observed when ballistic conduction occurs via edge states using the Landau-Büttiker formalism[2]. We present the calculation for two top-gates since the matrix size increases with increasing number of top-gates. The procedure can be generalized for $N$ top-gates, where $N$ is 37 in our device geometry. The schematic (Figure S5)(a) shows two top-gates and we assume equilibration among edge states. The assumption of equilibration is incorporated in our calculation using virtual voltage probe which enforces all the edge states to be at the same chemical potential in that terminal. The real voltage probes are marked in yellow. The gray probes denote virtual voltage terminal which enforces equilibration along the electrostatic edge of the top-gates. $|\nu_{bg}|$ is greater than $|\nu_{tg}|$ and the adjacent region have same polarity of charge carriers. By counting the number of current carrying channels that start from terminal $p$ and end in terminal $q$, transmission function $T_{pq}$ is obtained. The conductance matrix $G_{pq}$ is given by multiplying the following matrix with $e^2/h$,

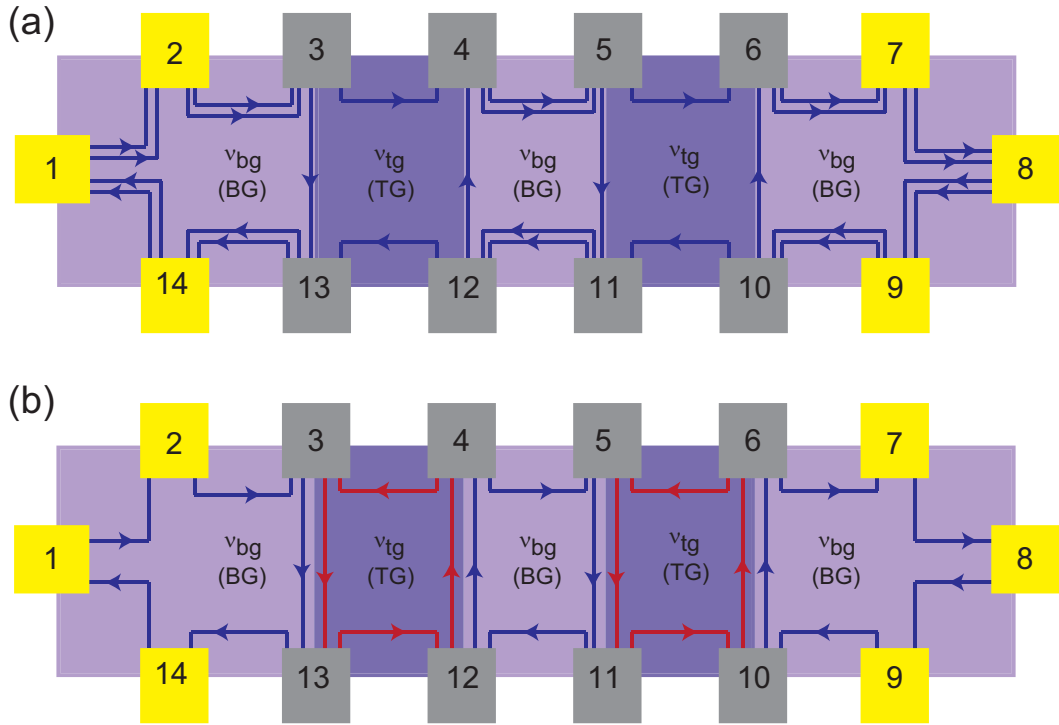

FIG. S5. Schematic showing two top-gates with the yellow probes as the actual contacts used to measure $R_{xx}$ and $R_{xy}$. Gray probes are the virtual voltage terminals used to enforce equilibration of the edge states along the electrostatic edge of the top-gate. (a) Equilibration when $|\nu_{bg}|$ is greater than $|\nu_{tg}|$ in the unipolar region. (b) Equilibration along all the electrostatic edges of the two top-gates, with $|\nu_{bg}|$ and $|\nu_{tg}|$ having opposite polarity in the bipolar region.

$$\begin{array}{c|cccccccccccccc}
\diagdown q & 1 & 2 & 3 & 4 & 5 & 6 & 7 & 8 & 9 & 10 & 11 & 12 & 13 & 14 \\
p & & & & & & & & & & & & & & \\
\hline
1 & 0 & 0 & 0 & 0 & 0 & 0 & 0 & 0 & 0 & 0 & 0 & 0 & 0 & |\nu_{bg}| \\
2 & |\nu_{bg}| & 0 & 0 & 0 & 0 & 0 & 0 & 0 & 0 & 0 & 0 & 0 & 0 & 0 \\
3 & 0 & |\nu_{bg}| & 0 & 0 & 0 & 0 & 0 & 0 & 0 & 0 & 0 & 0 & 0 & 0 \\
4 & 0 & 0 & |\nu_{tg}| & 0 & 0 & 0 & 0 & 0 & 0 & 0 & 0 & |\nu_{bg}|-|\nu_{tg}| & 0 & 0 \\
5 & 0 & 0 & 0 & |\nu_{bg}| & 0 & 0 & 0 & 0 & 0 & 0 & 0 & 0 & 0 & 0 \\
6 & 0 & 0 & 0 & 0 & |\nu_{tg}| & 0 & 0 & 0 & 0 & |\nu_{bg}|-|\nu_{tg}| & 0 & 0 & 0 & 0 \\
7 & 0 & 0 & 0 & 0 & 0 & |\nu_{bg}| & 0 & 0 & 0 & 0 & 0 & 0 & 0 & 0 \\
8 & 0 & 0 & 0 & 0 & 0 & 0 & |\nu_{bg}| & 0 & 0 & 0 & 0 & 0 & 0 & 0 \\
9 & 0 & 0 & 0 & 0 & 0 & 0 & 0 & |\nu_{bg}| & 0 & 0 & 0 & 0 & 0 & 0 \\
10 & 0 & 0 & 0 & 0 & 0 & 0 & 0 & 0 & |\nu_{bg}| & 0 & 0 & 0 & 0 & 0 \\
11 & 0 & 0 & 0 & 0 & |\nu_{bg}|-|\nu_{tg}| & 0 & 0 & 0 & 0 & |\nu_{tg}| & 0 & 0 & 0 & 0 \\
12 & 0 & 0 & 0 & 0 & 0 & 0 & 0 & 0 & 0 & 0 & |\nu_{bg}| & 0 & 0 & 0 \\
13 & 0 & 0 & |\nu_{bg}|-|\nu_{tg}| & 0 & 0 & 0 & 0 & 0 & 0 & 0 & 0 & |\nu_{tg}| & 0 & 0 \\
14 & 0 & 0 & 0 & 0 & 0 & 0 & 0 & 0 & 0 & 0 & 0 & 0 & |\nu_{bg}| & 0 \\
\end{array} \tag{2}$$

The current at a terminal $p$ is given by $I_p = \sum_q G_{pq}(V_p - V_q)$. All of these equations are not independent, so, voltage at the drain terminal (terminal 8) is chosen to be zero, and we omit the row and the column corresponding to this terminal. Inverting the resulting matrix, voltages at the terminals are obtained. Also, the current at the voltage terminals are zero.

The two probe voltage ($V_{2-probe}$) is given by $(V_1 - V_8)$, the four-probe longitudinal voltage ($V_{xx}$) is given by $(V_2 - V_7)$ and the Hall voltage ($V_{xy}$) is given by $(V_2 - V_{14})$. Dividing by current $I_1$, the resistances are obtained. Using the above formalism,

$$R_{2-probe} = \frac{e^2}{h}\frac{2|\nu_{bg}|-|\nu_{tg}|}{|\nu_{bg}||\nu_{tg}|}, \qquad R_{xx} = 2\frac{e^2}{h}\frac{|\nu_{bg}|-|\nu_{tg}|}{|\nu_{bg}||\nu_{tg}|}, \qquad R_{xy} = \frac{e^2}{h}\frac{1}{|\nu_{bg}|}. \tag{3}$$

The above calculation is done for two top-gates. We have numerically calculated resistances for N top-gates using the above formalism and obtained,

$$R_{2-probe} = \frac{e^2}{h}\frac{N|\nu_{bg}|-(N-1)|\nu_{tg}|}{|\nu_{bg}||\nu_{tg}|}, \qquad R_{xx} = N\frac{e^2}{h}\frac{|\nu_{bg}|-|\nu_{tg}|}{|\nu_{bg}||\nu_{tg}|}, \qquad R_{xy} = \frac{e^2}{h}\frac{1}{|\nu_{bg}|}. \tag{4}$$

Similarly, for $|\nu_{tg}|$ greater than $|\nu_{bg}|$ for N top-gates,

$$R_{2-probe} = \frac{e^2}{h}\frac{N|\nu_{tg}| - (N-1)|\nu_{bg}|}{|\nu_{bg}||\nu_{tg}|}, \qquad R_{xx} = N\frac{e^2}{h}\frac{|\nu_{tg}| - |\nu_{bg}|}{|\nu_{bg}||\nu_{tg}|}, \qquad R_{xy} = \frac{e^2}{h}\frac{1}{|\nu_{bg}|}; \quad (5)$$

The schematic for electron edge states and hole edge states in the adjacent region, that is, $|\nu_{tg}|$ and $|\nu_{bg}|$ having opposite polarity for two top-gates is shown in Figure S5(b). The virtual gray terminal ensures equilibration among electron and hole edge states. The above mentioned procedure can be followed to obtain the resistance as,

$$R_{2-probe} = \frac{e^2}{h}\frac{(N+1)|\nu_{tg}| + N|\nu_{bg}|}{|\nu_{bg}||\nu_{tg}|}, \qquad R_{xx} = N\frac{e^2}{h}\frac{|\nu_{tg}| + |\nu_{bg}|}{|\nu_{bg}||\nu_{tg}|}, \qquad R_{xy} = \frac{e^2}{h}\frac{1}{|\nu_{bg}|}, \quad (6)$$

where $N$ is the number of top-gates.

In the unipolar region, the observed $R_{xx}$ for 37 top-gates corresponds to the equilibration in case of a single top-gate. One possible reason is the narrow region in graphene defined electrostatically which facilitates the transmission of edge states along the length of graphene without equilibration with locally circulating edges. The schematic of this case is shown in Figure S6 where the gray probes along the extreme edges denote equilibration along those two particular edge. Another reason could be that the extreme edges of the top-gate are near ($\sim$ 300 nm) the actual voltage probe which helps in the equilibration along those edges.

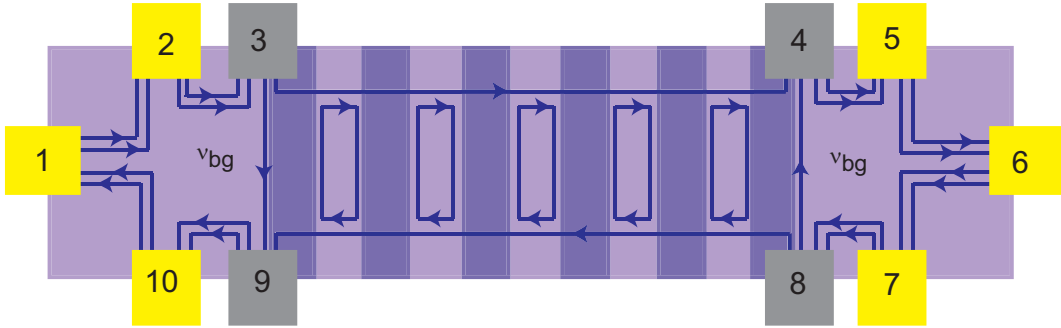

FIG. S6. Schematic with yellow probes as the real contacts and the gray probes as the virtual voltage terminals denoting equilibration of edge states along that edge. Schematic showing equilibration along the extreme edges of the top-gate in the unipolar region where $|\nu_{bg}|$ is greater than $|\nu_{bg}|$.

## VI. VARIATION OF HALL RESISTANCE WITH GATE VOLTAGES

Lack of equilibration or deviation from Landau-Büttiker picture is also observed in the Hall resistance ($R_{xy}$). In our device geometry, the electrodes measuring $R_{xy}$ is outside the array of finger gates and hence the edge state equilibration gives $R_{xy}$ to be a function of only $\nu_{bg}$ as seen in the previous section.

Figure S7(a) shows $R_{xy}$ measured as function of $V_{bg}$ and $V_{tg}$ at 14 T. A horizontal line slice showing plateaus at $\nu_{bg} = \pm 2, \pm 6$ is shown in Figure S7(b). Figure S7(c,d) shows R$_{xy}$ as function of $V_{tg}$ for $\nu_{bg} = 2, 6$. $R_{xy}$ being a function of only $\nu_{bg}$ should be independent of $V_{tg}$. We observe that when there are same type of charge carrier in the adjacent region, that is, for positive $\nu_{bg}$ (positive $V_{tg}$), $R_{xy}$ is almost constant. However, when we have electrons and holes in the adjacent region (positive $\nu_{bg}$ and negative $V_{tg}$), $R_{xy}$ does not remain constant. As we go from p-p' (or n-n') to p-n', $R_{xy}$, from its constant value shows a dip and then increases toward the constant value. This further supports our picture that there is enhanced scattering in the elements of the junction.

## VII. NUMERICAL SIMULATION OF ELECTROSTATICS USING COMSOL

We have modeled our device geometry in COMSOL software and obtained the electric field from numerical simulation of the electrostatics. In the simulation, we used 15 top-gates along the length of graphene and obtained the induced charge carrier density in graphene for a given back-gate and top-gate voltage. This induced charge carrier density is periodic along the length of graphene from which the potential is obtained. The electric field in the plane of graphene is obtained by taking the numerical derivative of the potential. The potential and the electric field variation along length of graphene for $V_{bg}$ = -30 V and $V_{tg}$ = 3 V in the bipolar region are shown in Figure S8(a,b). In the simulation, the geometrical width of the top-gate electrode is 30 nm, and the top-gates are 150 nm apart, similar to our device geometry. Due to the finite thickness of top-gate dielectric, there is a spread of the electric field lines from the top-gates. So, electrostatically, the effective width of the top-gated region in graphene is larger than 30 nm. For clarity, magnified images with 3 top-gates along length of graphene are shown in unipolar region (Figure S8(c)) and bipolar

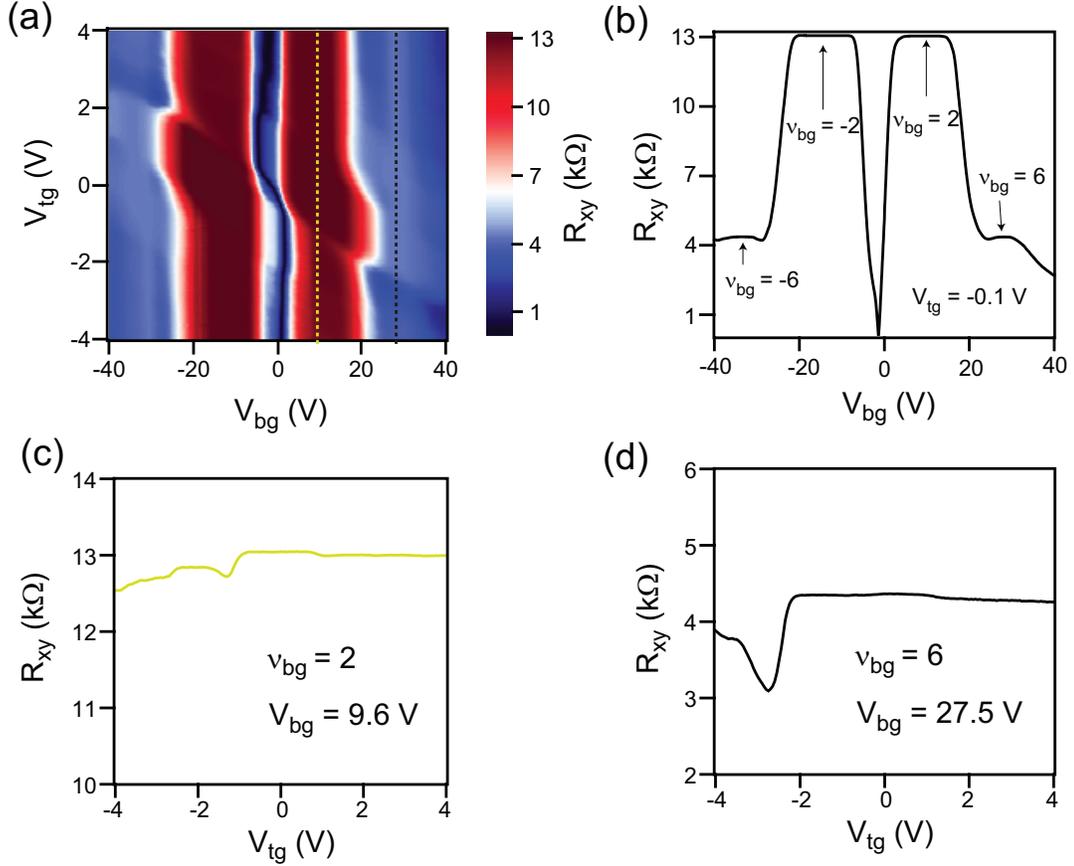

FIG. S7. Transverse resistance ($R_{xy}$) at 14 T at a temperature of 2 K. (a) Magnitude of $R_{xy}$ as a function of $V_{tg}$ and $V_{bg}$. (b) Magnitude of $R_{xy}$ as a function of $V_{bg}$ when $V_{tg}$ =-0.1 V. Plateaus at $\nu_{bg} = \pm 2$ and $\pm 6$ are observed. $R_{xy}$ as function of $V_{tg}$ at (c) $\nu_{bg} = 2$ and (d) $\nu_{bg} = 6$. In our device geometry, electrodes measuring $R_{xy}$ is outside the array of finger gates and hence $R_{xy}$ should be a function of only $\nu_{bg}$, that is, it should be constant with $V_{tg}$. However, $R_{xy}$ decreases from its constant value when $V_{tg}$ is such that there is opposite type of charge carrier in adjacent region.

region (Figure S8(d)). We observe the magnitude of electric field to be larger in bipolar region compared to the unipolar region. Another interesting thing to note is that not only the magnitude but also the shape of the potential is different in the two cases.

Numerical simulation of spatial variation of electric field along length and width of graphene in bipolar region ($V_{bg}$ = -30 V and $V_{tg}$ = 3 V) is shown in Figure S9(a). Electric field varies periodically along the length of graphene with maxima and minima at the edge of the top-gate, and is constant along the width of graphene. From the simulation, the max-

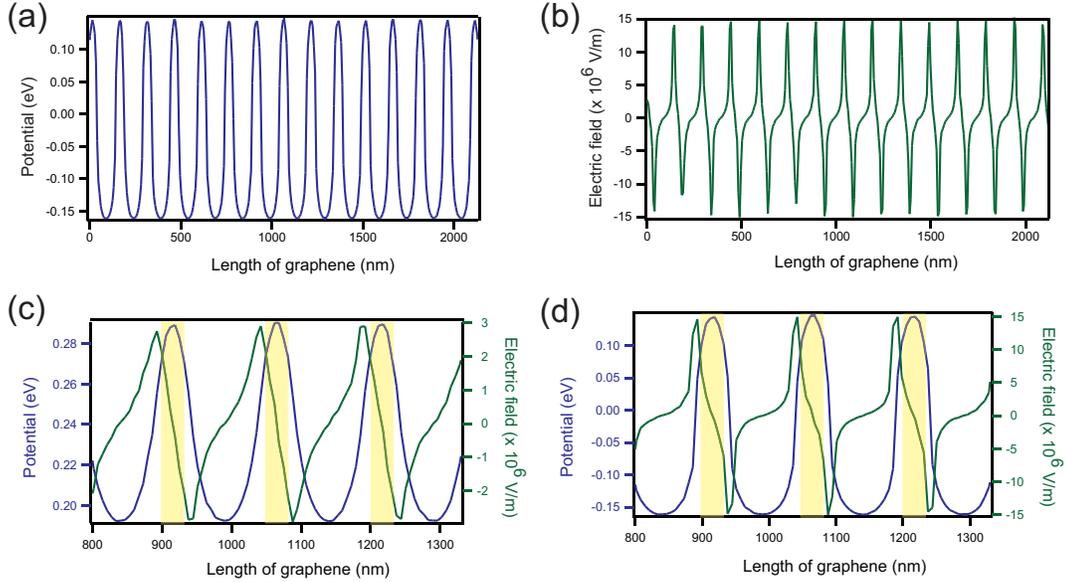

FIG. S8. Numerical simulation with 15 top-gates along length of graphene using COMSOL. (a) Variation of potential along length of graphene when $V_{tg}$ = 3 V and $V_{bg}$ = -30 V. (b) Spatial variation of electric field when $V_{tg}$ = 3 V and $V_{bg}$ = -30 V. Magnified image of spatial variation of potential and electric field in (c) unipolar region when $V_{tg}$ = 3 V and $V_{bg}$ = 30 V and (d) bipolar region when $V_{tg}$ = 3 V and $V_{bg}$ = -30 V. Yellow regions denote geometric position of top-gates.

imum electric field for a given $V_{bg}$ and $V_{tg}$ is obtained. Figure S9(b) shows the maximum value of electric field as a function of $V_{bg}$ and $V_{tg}$ in the bipolar region.

## VIII. EFFECT OF ELECTRIC FIELD ON EQUILIBRATION

As seen in the previous section, the potential and the electric field is high in bipolar region where we have a series of p-n' junctions compared to unipolar regime where there is series of n-n' (or p-p') junctions. In the unipolar region, there is lack of equilibration of edge states along the electrostatic edge and the array of top-gates behave as a single top-gate. The resistance plateaus depends only on the combination of filling factors, that is, resistance plateau at $(\nu_{tg},\nu_{bg})$ = (2,6) and (6,2) are same. In the bipolar region, the electric field modifies this lack of equilibration along the electrostatic edge such that the resistance plateau does not depend only on the combination of filling factor in the adjacent region. The electric field causes modification of Landau level wavefunction. With increasing electric field

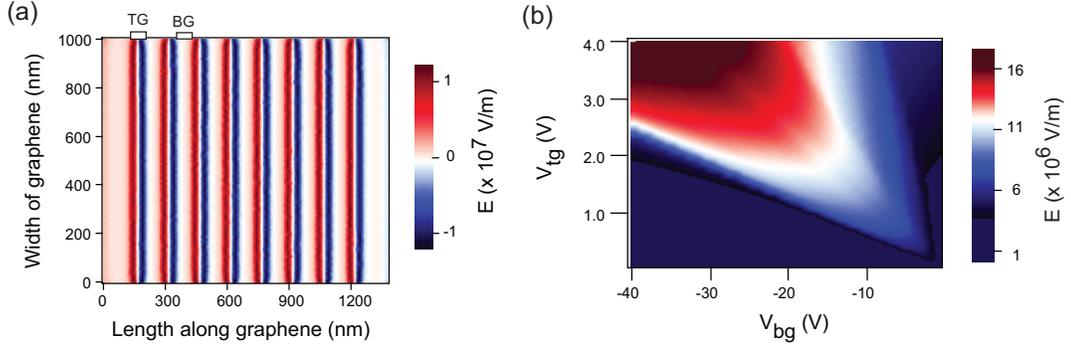

FIG. S9. Numerical simulation of electric field in our device in the bipolar region. (a) Spatial variation of the electric field along the length and width of graphene when $V_{tg} = 3$ V and $V_{bg} = -30$ V. The effective electrostatic widths of the TG and the BG regions for this gate voltage are marked in white. (b) Maximum electric field as function of $V_{tg}$ and $V_{bg}$.

the spread of the wavefunction increases such that it is of the order of the width of TG or BG region, resulting in enhanced scattering across Landau levels. This effect is manifested in charge transport when electric field becomes comparable to magnetic field times the Fermi velocity. Variation of effective magnetic length at 8 T as a function of gate voltages is shown in Figure S10.

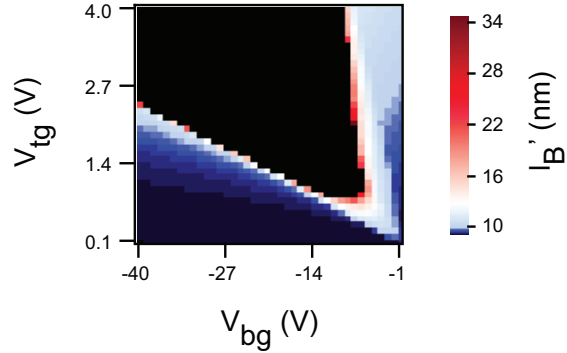

FIG. S10. Effective magnetic length at 8 T as a function of $V_{tg}$ and $V_{bg}$. The effective magnetic length is calculated from the maximum electric field obtained from the simulation as shown in the previous section.

In the bipolar region, the value of resistance depends on the electric and magnetic field in that region. For example, resistance value at $(\nu_{tg},\nu_{bg}) = (2,-6)$ and $(6,-2)$ are different as

the potential and the electric field are different in these two regions. Similarly resistance value for $(\nu_{tg},\nu_{bg}) = (6,-2)$ at 14 T and 8 T are different as shown in Figure S11(a,b).

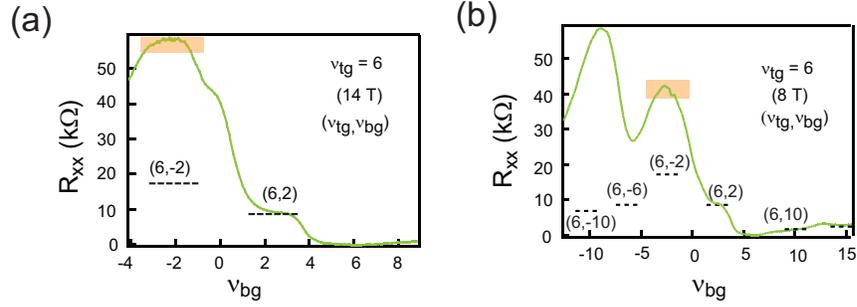

FIG. S11. Dependence of magnetotransport on electric field. $R_{xx}$ as function of $\nu_{bg}$ for $\nu_{tg} = 6$ at (a) 14 T and (b) 8 T. In the absence of equilibration and high electric field, resistance value for $(\nu_{tg},\nu_{bg}) = (6,-2)$ is different at different magnetic field.

## IX. EFFECT OF ELECTRIC FIELD ON MAGNETOTRANSPORT IN UNIPOLAR REGION

In the unipolar region, the electric field is low compared to the bipolar region. So, to observe the modification of Landau levels by electric field in the unipolar region, we measured $R_{xx}$ as a function of gate voltages from 2 T to 3.5 T. At a constant $V_{bg}$ of 20 V, $R_{xx}$ as a function of $V_{tg}$ shows SdH oscillations as magnetic field is increased above 3 T as seen in Figure S12(a).

We measured $R_{xx}$ as a function of $V_{tg}$ and $V_{bg}$ at 3.5 T (Figure S12(b)) in the unipolar region. Line plot of $R_{xx}$ vs $V_{bg}$ at $V_{tg} = 1$ V is shown by the green curve in Figure S12(c). At a constant $B$ of 3.5 T, we observe $R_{xx}$ to oscillate as function of $V_{bg}$ for a given $V_{tg}$. The oscillations arise due to periodic decrease of back-scattering as the Fermi energy moves through different Landau levels in BG region.

At $V_{tg} = 2.8$ V, the electric field is higher compared to $V_{tg} = 1$ V. So we took line slice of $R_{xx}$ vs $V_{bg}$ at $V_{tg} = 2.8$ V (orange curve in Figure S12(c)) and observed the fading of SdH oscillations. So with increasing electric field, the Landau levels come close to each other resulting in a decrease in the modulation in resistance. Thus the decrease in amplitude of oscillation is consistent with the modification of Landau levels with increasing electric field.

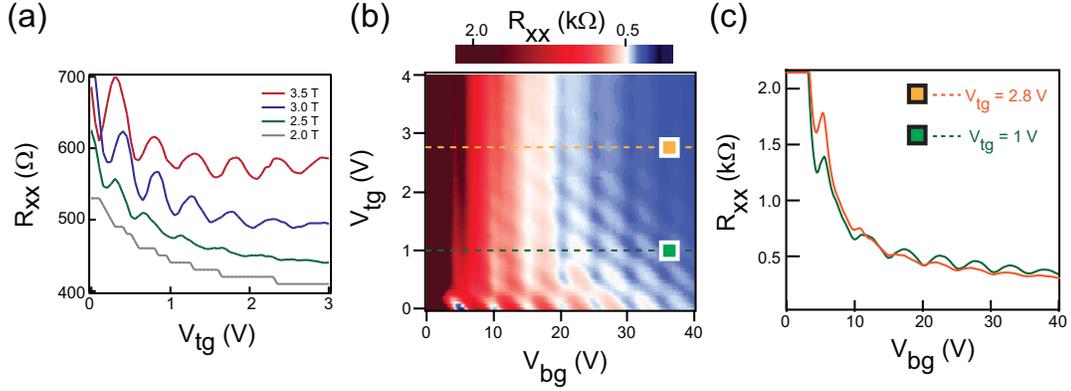

FIG. S12. Effect of electric field on resistance oscillations in unipolar region (a) $R_{xx}$ as a function of $V_{tg}$ at $V_{bg} = 20$ V at different magnetic field. SdH oscillations are observed when magnetic field is increased above 3 T. The plots at different magnetic fields are offset for clarity. (b) $R_{xx}$ as a function of $V_{tg}$ and $V_{bg}$ at 3.5 T in the unipolar region. (c) Line plot of $R_{xx}$ as a function of $V_{bg}$ at $V_{tg} = 1$ V (green curve) and 2.8 V (orange curve). Amplitude of oscillations decreases with increasing $V_{tg}$ and hence increasing electric field.

## X. MEASUREMENTS FROM ANOTHER DEVICE

In this section we present measurements of another device at 14 T which showed similar magnetotransport behavior. This device has a low mobility of 2009 cm$^2$/Vs, and the ratio of top-gate capacitance to back-gate capacitance is 11. The number of top-gates in this device is 35 and the period is 150 nm.

Figure S13(a) shows the variation of $R_{xx}$ as a function of $V_{tg}$ and $V_{bg}$ at 14 T. All the plateaus are not well resolved due to low mobility. Dissipationless transport is observed for $(\nu_{tg},\nu_{bg}) = (2,2)$, (-2,-2) and (-6,-6). The maximum resistance in this device, observed in the bipolar region, is 58 k$\Omega$.

Figure S13(b) shows line slice of $R_{xx}$ as a function of $\nu_{bg}$ at $\nu_{tg}$ of -2. The black dashed lines correspond to edge state equilibration for a single top-gate. The experimental curve (in green) coincides with the black dashed lines at $(\nu_{tg},\nu_{bg}) = (-2,-6)$. Thus this device having 35 top-gates shows the plateau in $R_{xx}$ corresponding to a single top-gate in the unipolar region.

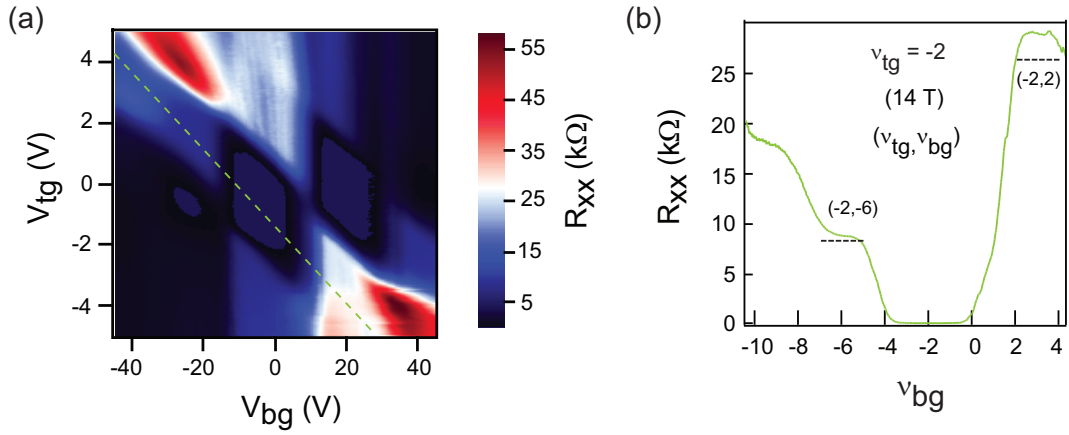

FIG. S13. Charge transport measurement of another device in the presence of a magnetic field of 14 T. (a) $R_{xx}$ as a function of $V_{tg}$ and $V_{bg}$. (b) $R_{xx}$ as a function of $\nu_{bg}$ for $\nu_{tg} = -2$. Resistance plateaus at $(\nu_{tg},\nu_{bg})$ = (-2,-6) and (-2,2) for a single top-gate are marked by black dashed lines.



\end{document}